\documentclass[pre]{revtex4}
\usepackage{amsmath}
\usepackage{enumerate}
\usepackage{graphicx}

\begin{document}

\title{Longitudinal Oscillations in a Nonextensive Relativistic Plasma}
\author{V\'{\i}ctor Mu\~noz}
\affiliation{Department of Earth System Science and Technology, Kyushu
  University, Fukuoka 816-8580, Japan}
\affiliation{Departamento de F\'{\i}sica, Facultad de Ciencias,
Universidad de Chile, Casilla 653, Santiago, Chile}
\begin{abstract}
  
  The dispersion relation of longitudinal electrostatic oscillations
  in a relativistic plasma is studied in the context of the
  nonextensive statistics formalism proposed by Tsallis [C. Tsallis,
  J.~Stat.~Phys.~{\bf 52}, 479 (1988)], where nonextensivity is
  characterized by a parameter $q$ in Tsallis's entropy. $q=1$ corresponds to
  the usual Boltzmann-Gibbs, extensive statistics formalism. In the
  nonrelativistic regime, normalizability of the equilibrium
  distribution function implies that $-1\leq q\leq\infty$. We show that in the
  relativistic regime much tighter constraints must be satisfied,
  namely $0\leq q \leq 1+ k_B T/mc^2$, where $k_B$ is the Boltzmann
  constant, $T$ is the temperature of the
  plasma, and $m$ is the particle mass.
 
  Then we study longitudinal oscillations in a proton-electron plasma,
  assuming immobile protons, and electrons whose distribution function
  maximizes Tsallis's entropy. The dispersion relation of these
  oscillations is written in integral form for the long wavelength
  limit. Explicit expressions in terms of generalized hypergeometric
  functions can be found for all possibles values of $q$ in the
  ultra-relativistic regime.
\end{abstract}
\maketitle

\section{Introduction}

Traditionally, the equilibrium of statistical systems has been studied
based on the Boltzmann-Gibbs entropy, 
\begin{equation}
\label{bg}
  S_{BG} = -k_B \sum_i p_i \ln p_i \ ,
\end{equation}
where $k_B$ is the Boltzmann constant, and $p_i$ is the probability of
the $i$-th microstate. However, an increasing amount of experimental,
computational and theoretical evidence, shows that this formalism
is not adequate for systems where long range
interactions in time or space are important, and, in general, systems where the
relevant phase space has fractal or multifractal structure. Tsallis
\cite{Tsallis} proposed the following form for the entropy of such systems:
\begin{equation}
  \label{tsallis}
  S_q = k_B \frac{1-\sum_i p_i^q}{q-1} \ ,
\end{equation}
where $q$ is a real number. For $q=1$, $S_q$ reduces to $S_{BG}$. 

$S_q$ has the basic property that for independent systems $A$ and $B$,
the entropy of the composite system $A+B$ is given by:
\begin{equation}
  \label{nonextensive}
  S_q(A+B) = S_q(A) + S_q(B) + (1-q)S_q(A)S_q(B) \ . 
\end{equation}
Thus, $q$ is a measure of the nonextensivity. Among the various
physical systems where connections with the Tsallis entropy formalism
have been found are gravitational
systems \cite{Taruya}, long range Hamiltonian systems \cite{Latora},
nonlinear inverse bremsstrahlung absorption in plasmas
\cite{Tsallis_a}, turbulence \cite{Arimitsu}, and granular systems
\cite{Satin}, and many applications in astrophysics, fluid
dynamics, biology, economy, quantum mechanics, and others. (See,
e.g. \cite{Abe}.)

An important property of the nonextensive formalism is that the
distribution function which maximizes $S_q$ is
non-Maxwellian \cite{Silva_a,Lima}. Specifically, for $q<1$, high
energy states are more probable than in the extensive case; for
$q>1$ high energy states are less probable than in the extensive case,
and there is a cutoff beyond which no states exist. For a
unidimensional gas where
$E=mv^2/2$, this cutoff is given by \cite{Lima} 
\begin{equation}
  \label{cutoff_nonrelativistic}
  v_\text{max} = \sqrt{2k_BT/m(q-1)} \ .
\end{equation}

Velocity distributions in plasmas are often found to be
non-Maxwellian. High energy tails in space and laboratory plasmas
\cite{Mauk,Scudder,Marsch_a,Collier, Liu_a} are a common occurrence.
Metaequilibrium non-Maxwellian radial density profiles in cylindrical
turbulent pure electron plasmas are observed \cite{Huang}. As stated
above, such
distribution functions emerge naturally if the system is described by
a nonextensive statistics, and several
authors have successfully applied this approach to plasma systems
\cite{Boghosian,Anteneodo,Tsallis_a,Leubner}. 

Lima et al.~\cite{Lima} studied the longitudinal oscillations of a
collisionless thermal plasma, in the context of the nonextensive
formalism based on \eqref{tsallis}. They derived the dispersion
relation for electrostatic waves, and calculated it explicitly for an
equilibrium distribution function which maximizes
\eqref{tsallis}. They found that it fits experimental data better than
if the energy distribution is given by a Maxwellian.  

All the previous analysis have been made for non-relativistic
plasmas. However, relativistic extensions of the Tsallis formalism are
possible \cite{Lavagno}. Our purpose is to study the consequences of
a relativistic nonextensive statistics in plasmas. In particular, we
will examine some general properties of the equilibrium distribution function
and calculate the dispersion relation for electrostatic waves in the
long wavelength limit for ultrarelativistic plasmas.



\section{Relativistic distribution function}
\label{distribution_function}

In the nonextensive description, the equilibrium distribution function
for a relativistic plasma can be written \cite{Lavagno}
\begin{equation}
  \label{fq}
  f_q(\vec p\,) = A_q \left[ 1 - (q-1) \frac E{k_B
      T} \right]^{1/(q-1)} \ , 
\end{equation}
where 
\begin{equation}
  \label{energy}
  E = \sqrt{m^2 c^4 + \vec p\,^2 c^2} \ ,
\end{equation}
is the particle energy and $A_q$ is a normalization constant.

Equation \eqref{fq} generalizes the nonrelativistic expression (see,
e.g., \cite{Lima}). For $q=1$, \eqref{fq} yields the usual Boltzmann
distribution function $f_{q=1} = A_1 \exp(-E/k_B T)$.

For simplicity, in the following 
we will consider a one dimensional plasma, so that
$\vec p = p\hat z$.

Since
the relativistic energy is positive, $f_q$ will be real and
normalizable if $E$ satisfies the following conditions:
\begin{align}
\label{q_gt_1}
  \text{if $q\geq1$} & \qquad 0\leq E\leq\frac{k_BT}{q-1} \ , \\
\label{q_lt_1}
  \text{if $q\leq1$} & \qquad 0\leq E \ ,
\end{align}
i.e., $E$ can take any possible value if $q\leq1$, but the distribution
function has an energy cutoff if $q\geq1$. Notice that both \eqref{q_gt_1} and
\eqref{q_lt_1} hold for the extensive limit $q=1$, in which case they
give the expected constraint $0\leq E < \infty$.

Due to the cutoff
\eqref{q_gt_1}, if $q\geq1$ $f_q$ is always normalizable. 
However, normalizability of the distribution function for $0\leq E
\leq \infty$
when $q < 1$ is not guaranteed.
 The integral 
\begin{equation}
\label{integral}
  I = \int_{-\infty}^\infty f_q(p) \, dp \ ,
\end{equation}
is finite if $f_q(p) \sim p^s$ when $p \to\infty$, with $s<-1$.
 Using \eqref{energy}, 
 \begin{equation*}
   f_q(p) \xrightarrow[p\to\infty]{} C  p^{\frac{1}{q-1}} \ ,
 \end{equation*}
where $C=\left[-\frac{q-1}{k_BT}\right]^{\frac{1}{q-1}}>0$. Thus,
$f_q(p)$ is normalizable if $q>0$.


Let us now analyze the case $q\geq 1$. The cutoff in \eqref{q_gt_1}
and \eqref{energy} give
\begin{equation}
  \label{cutoff_p}
  |p| \leq \frac 1c \sqrt{\left( \frac{k_BT}{(q-1)}\right)^2 - m^2c^4
    } \ .
\end{equation}
If 
\begin{equation}
\label{condicion_p_real}
k_BT/(q-1)\geq mc^2 \ ,
\end{equation}
then \eqref{cutoff_p} yields the cutoff in
momentum space corresponding to the energy cutoff \eqref{q_gt_1}. 
In the limit case 
$k_BT/(q-1)= mc^2$, the only momentum allowed is $p=0$. 
If $k_BT/(q-1)< mc^2$, then no real value of $p$ satisfies 
condition \eqref{q_gt_1}, and no distribution function
exists. We may interpret this saying that for a given temperature, only 
$q$ values consistent with \eqref{condicion_p_real} are possible:
\begin{equation}
  \label{condicion_q}
  q \leq 1+ \frac{k_BT}{mc^2} \ . 
\end{equation}

\clearpage

Summarizing:

\begin{enumerate}[\itshape {Case } 1:] 
\item $0\leq q$.

Forbidden $q$. $f_q$ is not normalizable.

\item $0<q\leq 1$.

$f_q$ is normalizable. Energy can take values 
$$0\leq E <\infty \ .$$

\item $1\leq q \leq 1 + \dfrac{k_BT}{mc^2}$.

$f_q$ is normalizable. Energy can only take values 
$$0\leq E\leq \dfrac{k_BT}{q-1}\ ,$$
i.e.
$$0 \leq |p| \leq \frac 1c \sqrt{\left( \frac{k_BT}{(q-1)}\right)^2 - m^2c^4
    } \ .$$

\item $1 + \dfrac{k_BT}{mc^2} < q$.

Forbidden $q$. No values of $p$ consistent with energy cutoff.
\end{enumerate}

Two main differences arise with respect to the 
 non-relativistic case \cite{Lima},
where allowed values are $-1\leq q < \infty$. In the relativistic case 
 constraints are much tighter, $0\leq q \leq 1+
 k_BT/mc^2$, and  depend on the ratio of thermal to rest energy.

\vspace{.3cm}

Considering these constraints, the relativistic distribution function
can be plotted. The result is qualitatively similar to the
nonrelativistic case:

\vspace{.5cm}
\begin{figure}[h]
\includegraphics[width=.4\textwidth]{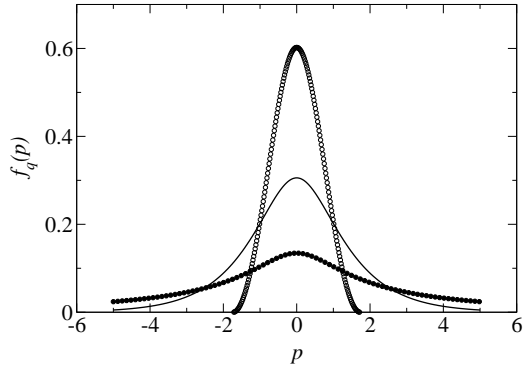}
\caption{Relativistic dispersion function \eqref{fq}
  ($k_BT=m=c=1$). Full circles: 
  $q=0.5$. Solid line: $q=1$. Open circles: $q=1.5$.}
\end{figure}


\section{Dispersion relation for longitudinal plasma oscillations}

We consider a proton-electron relativistic plasma. Assuming protons remain
at rest, and that particles only move in one dimension ($\hat z$),
the dispersion relation for longitudinal oscillations is
\begin{equation}
  \label{disprel}
  1 = i\frac{\omega_p^2}k m  \int dp_z\,  \frac{\partial
    f_q(p_z)/\partial p_z}{s+i k v_z} \ ,
\end{equation}
where $\omega_p$ is the plasma frequency, $k$ is the wavenumber,
$s$ is the argument of the Laplace transform ($s=-i\omega$, where
$\omega$ is the wave frequency, if there is no damping), 
$m$ the electron mass, $p = m\gamma
v$ is the particle momentum, and  $f_q(p,z)$ is
the equilibrium electron distribution function, normalized to unity.
Particle energy is  $E= \sqrt{m^2c^4 + p_z^2c^2}$.

Following Ref.~\cite{Lima}, we 
now consider the long wavelength limit $k\ll k_D$, where $k_D
= 2\pi/\lambda_D$, and $\lambda_D$ is the Debye length. The integrand
has a pole at $v=-is/k$. If $\mbox{Im } v\ll 1$, integration can be
performed on the real axis:
\begin{equation}
  \label{disprel_Taylor}
  1 \simeq \frac{im\omega_p^2}{sk}
  \int_{-p_{\text{max}}}^{p_{\text{max}}} dp_z \, \frac{\partial
    f_q}{\partial p_z} \left( 1 - i\frac{kv_z}{s} + \cdots \right) \ ,
\end{equation}
where $p_{\text{max}}$ is the maximum momentum allowed by
\eqref{cutoff_p} if $1\leq q \leq 1 + {k_BT}/{mc^2}$, or $\infty$ if
$0<q\leq1$. Since $f_q$ is an even function of $p_z$, the first term
in \eqref{disprel_Taylor} vanishes. The dispersion relation can then 
 be written:
\begin{equation*}
  1 = \frac{m\omega_p^2}{s^2} \int_{-p_\text{max}}^{p_\text{max}}
  dp_z \, \frac{p_z}{m\gamma} \frac{\partial 
    f_q}{\partial p_z} \ ,
\end{equation*}
where $\gamma = (1+p_z^2/m^2 c^2)^{1/2}$. Integrating by parts, and
using the fact that $ f_q(\pm p_\text{max}) = 0$, 
finally yields 
\begin{equation}
\label{disprel_long_k}
  1 = -\frac{\omega_p^2}{s^2} \int_{-p_\text{max}}^{p_\text{max}}
  dp_z\, \frac{ f_q(p_z)}{\gamma^3} \ .
\end{equation}

Equation \eqref{disprel_long_k} depends on $q$ in general, and
therefore, unlike the nonrelativistic case \cite{Lima}, nonextensive
corrections to the dispersion relation of longitudinal waves
 may appear to the lowest order in $kv_z/s$.

\section{Ultrarelativistic plasma}

Closed expressions in terms of known functions can be obtained for
Eq.~\eqref{disprel_long_k} in the ultrarelativistic case $|p_z|\gg
mc$.
As shown in
Sec.~\ref{distribution_function}, $q$ can only take values between 0
and $1+ k_BT/mc^2$. We consider two situations:

\subsection{$0<q\leq1$}

Since $p_{\rm
  max}=\infty$, in this case, the dispersion relation
\eqref{disprel_long_k} can be written 
\begin{equation}
  \label{disprel_ultra}
  1 = -2\frac{\omega_p^2}{s^2} A_q \int_0^\infty dp_z\, \frac
  1{\left(1+\frac{p_z^2}{(mc)^2}\right)^{3/2}} \frac 1{\left(1 -(q-1) 
  \frac{p_zc}{k_BT}\right)^{-\frac1{q-1}}} \ . 
\end{equation}
Noting that $-(q-1)>0$ and $-1/(q-1)>0$, normalization of the
distribution function
\begin{equation}
  \label{fq_ultra_q}
 f_q(p_z) = A_q \frac 1{\left( 1-(q-1)
 \dfrac{|p_z|c}{k_BT}\right)^{-\frac{1}{q-1}}} \ ,
\end{equation}
\begin{equation}
  \label{normalization}
  1 = \int_0^\infty \hat f_q(p_z)\, dp_z \ ,
\end{equation}
yields
\begin{equation}
  \label{A_q_ultra}
  A_q = \frac{1}{mc}\frac q{2\tau} \ ,
\end{equation}
where
\begin{equation}
  \label{tau}
  \tau = \frac{k_BT}{mc^2} \ .
\end{equation}

The integral in \eqref{disprel_ultra} can be done analytically,
yielding
\begin{equation}
\label{disprel_ultra_I}
  1 = -\frac{\omega_p^2}{s^2} \frac 1\tau q I\left(\tau,-\frac 1{q-1}
    \right)
    \ , 
\end{equation}
where 
\begin{align}
  \label{I}
  I(\tau,\alpha) &= -\frac1{\sqrt{\pi}}
  (-\tau\alpha)^{\alpha+1}\alpha \,
  \Gamma\left(\frac{3+\alpha}2\right) \Gamma\left(-\frac \alpha2\right)
  \mbox{}_2F_1\left( \frac{1+\alpha}2,\frac{3+\alpha}2,\frac32;-
\alpha^2\tau^2\right) \nonumber \\
&+ \frac 1{\sqrt{\pi}} (-\tau\alpha)^\alpha\,
  \Gamma\left(\frac{1-\alpha}{2}\right) \Gamma\left(\frac{2+\alpha}2\right)
   \mbox{}_2F_1\left(
    \frac{2+\alpha}2,\frac{\alpha}2,\frac12;-\alpha^2\tau^2\right) \nonumber \\
&+  
  \frac{\tau\alpha}{1-\alpha} \mbox{}_3F_2 \left[ \left\{\frac 12,1,\frac 32\right\},
    \left\{\frac{2-\alpha}2,\frac{3-\alpha}2\right\},
  -\tau^2\alpha^2\right] \ ,
\end{align}
$\mbox{}_2F_1$ is the hypergeometric function,
\begin{equation}
  \label{2F1}
  \mbox{}_2F_1(a,b,c;z) = \sum_{\nu=0}^\infty \frac{(a)_\nu
    (b)_\nu}{(c)_\nu} \frac{z^\nu}{\nu!} \ , 
\end{equation}
with $(a)_\nu = a(a+1)\cdots (a+\nu-1)$, and $\mbox{}_mF_n$ is the
generalized hypergeometric function,
\begin{equation}
  \label{mFn}
  \mbox{}_mF_n(\{a_1,\ldots,a_m\},\{b_1,\ldots,b_n\},z) 
= \sum_{\nu=0}^\infty \frac{(a_1)_\nu \cdots
  (a_m)_\nu}{(b_1)_\nu\cdots (b_n)_\nu}\frac{z^\nu}{\nu!} \ .
\end{equation}



\subsection{$1\leq q \leq 1+ k_BT/mc^2$}

Now $p_{\rm max}$ is given by Eq.~\eqref{cutoff_p}, and dispersion
relation Eq.~\eqref{disprel_long_k} is

\begin{equation}
  \label{disprel_ultra_B}
  1 = -2\frac{\omega_p^2}{s^2} A_q \int_0^{p_{\rm max}} dp_z\, \frac
  1{\left(1+\frac{p_z^2}{(mc)^2}\right)^{3/2}} \left(1 -(q-1) 
  \frac{p_zc}{k_BT}\right)^{\frac1{q-1}} \ . 
\end{equation}
Noting that $q-1>0$, normalization of the distribution function yields
\begin{equation}
  \label{A_q_ultra_B}
  A_q = \frac q{2mc}\frac 1\tau \left[ 1-\left( 1-(q-1)\frac{\hat
        p}\tau \right)^{\frac q{q-1}}\right]^{-1} \ ,
\end{equation}
where 
\begin{equation}
  \label{p_hat}
  \hat p=\frac{p_{\rm max}}{mc} \ .
\end{equation}
Again, integration in \eqref{disprel_ultra_B} can be performed
analytically, yielding
\begin{equation}
\label{disprel_ultra_J}
  1 = -\frac{\omega_p^2}{s^2} \frac 1{\sqrt{1+\hat p^2}} 
J\left(\tau,\frac 1{q-1}\right) \ , 
\end{equation}
where
\begin{multline}
  \label{J}
  J(\tau,\alpha) = \left[ 1-\left(1-\frac{\hat p}{\alpha\tau}
    \right)^{\alpha+1} \right]^{-1} \frac 1{(\alpha^2\tau^2+1)^2} \\
\times \left[ (1+i\alpha\tau)\sqrt{1+\hat p^2}
\cdot 
F_1\left(1+\alpha;\frac32,\frac32;2+\alpha;\frac{\alpha\tau}{\alpha\tau+i}
,\frac{\alpha\tau}{\alpha\tau-i}\right)
\right. \\
\left. - \left(1-\frac{\hat p}{\alpha\tau}\right)^{1+\alpha}
[(1+i\alpha\tau)+\hat p(i+\alpha\tau)] \cdot
F_1\left(1+\alpha;\frac32,\frac32;2+\alpha;
\frac{\alpha\tau-\hat p}{\alpha\tau+i},
\frac{\alpha\tau-\hat p}{\alpha\tau-i}\right) \right] \ ,
\end{multline}
where $F_1$ is the 
Appell hypergeometric function of two variables:
\begin{equation}
\label{Appell}
  F_1(a;b_1,b_2;c;x,y) = \sum_{m=0}^\infty \sum_{n=0}^\infty
  \frac{(a)_{m+n} (b_1)_m (b_2)_n}{m!\, n!\, (c)_{m+n}} x^m y^n \ .
\end{equation}

\section{Conclusions}

Some consequences of describing a plasma based on a nonextensive
statistical treatment have been discussed. In particular, the one
dimensional plasma
distribution function which maximizes Tsallis's entropy $S_q$ has been
considered. In the nonrelativistic case, normalizability of the
distribution function constrains $q$ to values greater than or equal
to~$-1$. However, in the relativistic case a much shorter range is
possible, and the maximum value of $q$ depends on the ratio of
temperature to particle rest energy.

Then we derive the dispersion relation for longitudinal oscillations
in a plasma composed of electrons and rest protons. In the long
wavelength limit, and unlike the nonrelativistic case, nonextensive
corrections appear to the lowest order. Finally, in the ultra-relativistic
regime, this dispersion relation is written explicitly in terms of
generalized hypergeometric functions.


\end{document}